\documentclass[prb,reprint,amsmath,amssymb,aps]{revtex4-2}

\usepackage{graphicx}
\usepackage{dcolumn}
\usepackage{bm}

\usepackage[utf8]{inputenc}
\usepackage[T1]{fontenc}
\usepackage{mathptmx}

\begin{document}
\title{Transition to strong coupling regime in hybrid plasmonic systems: Exciton-induced transparency and Fano interference}
\author{Tigran V. Shahbazyan}
\affiliation{
Department of Physics, Jackson State University, Jackson, MS
39217 USA}


\begin{abstract} 
We present a microscopic model describing the transition to strong coupling regime for an emitter resonantly coupled to a surface plasmon in a metal-dielectric structure. We demonstrate that the  shape of scattering spectra is determined by an interplay of two distinct mechanisms. First is the near-field coupling between the emitter and the plasmon mode which underpins energy exchange between the system components  and gives rise to exciton-induced transparency minimum in scattering spectra prior the transition to strong coupling regime. The second mechanism is Fano interference between the plasmon dipole and the plasmon-induced emitter's dipole as the system interacts with the radiation field. We show that the Fano interference can strongly affect the overall shape of scattering spectra,  leading to the inversion of spectral asymmetry that was recently reported in the experiment.
\end{abstract}
\maketitle

\section{Introduction}
\label{sec:intro}

Strong coupling between surface plasmons in metal-dielectric structures  and excitons in semiconductors or dye molecules has recently attracted intense interest driven to a large extent by  possible applications in ultrafast reversible switching \cite{ebbesen-prl11,bachelot-nl13,zheng-nl16}, quantum computing \cite{waks-nnano16,senellart-nnano17}, and light harvesting \cite{leggett-nl16}. In the strong coupling regime, coherent energy echhange between excitons and plasmons  \cite{shahbazyan-nl19} leads to the emergence of mixed polaritonic states with energy bands separated by the anticrossing gap (Rabi splitting) \cite{novotny-book}. For excitons coupled to  cavity modes in  microcavities, the Rabi splitting magnitudes  are  relatively small on the scale of several meV  \cite{forchel-nature04,khitrova-nphys06,imamoglu-nature06}. However,   in hybrid plasmonic systems, where surface plasmons  are coupled to excitons in J-aggregates \cite{bellessa-prl04,sugawara-prl06,wurtz-nl07,fofang-nl08,lienau-prl08,bellessa-prb09,schlather-nl13,lienau-acsnano14,shegai-prl15,shegai-nl17,shegai-acsphot19}, in various dye molecules \cite{hakala-prl09,berrier-acsnano11,salomon-prl12,luca-apl14,noginov-oe16}  or in semiconductor nanostructures \cite{vasa-prl08,gomez-nl10,gomez-jpcb13,manjavacas-nl11}, the  Rabi splittings can be much greater even reaching  hundreds  meV. For single excitons, however, achieving  a strong exciton-plasmon coupling is a  challenging task as it requires extremely small plasmon mode volumes, which can mainly be achieved in nanogaps \cite{hecht-sci-adv19,pelton-sci-adv19,baumberg-natmat2019}.

At the same time, the scattering spectra of hybrid plasmonic systems, such as excitons in J-aggregates or colloidal QDs coupled to gap plasmons in nanoparticle-on-metal (NoM) systems \cite{haran-nc16,baumberg-nature16,lienau-acsph18,pelton-nc18} or those in two-dimensional atomic crystals conjugated with Ag or Au nanostructures \cite{zhang-nl17,xu-nl17,alu-oe18,urbaszek-nc18,shegai-nl18,zhang-acsphot19}, exhibit a narrow minimum even before reaching the strong coupling transition point. The emergence of such a minimum in the weak coupling regime is referred to as exciton-induced transparency (ExIT)  \cite{vuckovic-prl06,bryant-prb10,pelton-oe10}, in analogy to electromagnetically-induced transparency (EIT) in pumped three-level atomic systems that is attributed to  Fano interference between different excitation pathways. Recently, we have shown that, in the linear regime (i.e., in the absence of pump), the emergence of this minimum is due to imbalance of energy exchange between the emitter and plasmon in a narrow frequency interval \cite{shahbazyan-prb20}. Typically, the plasmon plasmon optical dipole moment significantly (by $\sim 10^{4}$)  exceeds that of an exciton in a semiconductor quantum dot  and so the emitter's direct interaction with the radiation field is relatively weak \cite{pelton-ns19}. In this case, the ExIT minimum in  scattering spectra is described, with a reasonably good accuracy, by the dressed plasmon model or by its classical analogue -- the coupled oscillators model, in which only the plasmon interacts with the radiation field, so that the scattering spectra show a narrow ExIT minimum on top of a broad plasmon band, while the overall spectral weight is tilted towards the higher frequency range \cite{pelton-oe10,pelton-nc18,shahbazyan-prb20}.

On the other hand, in hybrid plasmonic systems, the \textit{optical} interference between an exciton and a plasmon can arise from indirect coupling of exciton to the radiation field. Namely, if the incident light frequency is tuned to the plasmon resonance, the exciton dipole moment \textit{induced} by the plasmon near field  is not necessarily small, so that the exciton can substantially contribute, albeit indirectly, to  the system optical transition. This gives rise to  Fano interference between the plasmon and plasmon-induced exciton dipoles which can significantly affect the overall shape of optical spectra.  As we show in this paper, such  Fano interference effects can lead to \textit{ inversion of spectral asymmetry}, characterized by spectral weight shift towards lower frequency range, which was observed for excitons coupled to localized plasmon modes \cite{zhang-nl17,xu-nl17,shegai-acsphot19}.

In this paper, we present a microscopic model for linear optical response of a single exciton resonantly coupled to a surface plasmon mode in a metal-dielectric structure which accounts for both ExIT and Fano interference effects as the system transitions to strong coupling regime. Starting with the canonical Hamiltonian with microscopic coupling parameters \cite{shahbazyan-prb21}, we set up the system of Maxwell-Bloch equations for induced dipole moments which determine scattering spectrum of the hybrid plasmonic system. We further show that while the ExIT minimum results from the energy exchange imbalance in a narrow frequency interval, the overall spectral shape of scattering spectra is strongly affected by the Fano interference between radiating plasmon and plasmon-induced exciton dipoles. Specifically, we demonstrate that Fano interference can lead to an inversion of spectral asymmetry, consistent with the experiment \cite{zhang-nl17,xu-nl17,shegai-acsphot19}.

\section{The system Hamiltonian and microscopic coupling parameters}

We consider a quantum emitter (QE) with dipole moment $\bm{\mu}_{e}$ and excitation frequency $\omega_{e}$ situated at a position $\bm{r}_{e}$  near a metal-dielectric structure characterized by complex dielectric function  $\varepsilon (\omega,\bm{r})=\varepsilon' (\omega,\bm{r})+i\varepsilon'' (\omega,\bm{r})$ supporting localized plasmon modes with frequencies $\omega_{m}$ interacting with external  electromagnetic (EM) field $\bm{E}(t)$. For monochromatic EM field of frequency $\omega$, in the rotating wave approximation (RWA), the system dynamics is described by the Hamiltonian 
\begin{align}
\label{H-full}
H=
&\hbar\omega_{m}\hat{a}^{\dagger}\hat{a}+\hbar\omega_{e} \hat{\sigma}^{\dagger}\hat{\sigma}+\hbar g(\hat{\sigma}^{\dagger}\hat{a}+\hat{a}^{\dagger}\hat{\sigma})
\nonumber\\
&-\left (\bm{\mu}_{m}\!\cdot\!\bm{E}\,\hat{a}^{\dagger} e^{-i\omega t} + \bm{\mu}_{e}\!\cdot\!\bm{E}\,\hat{\sigma}^{\dagger} e^{-i\omega t} +  {\rm H.c.}\right ),
\end{align}
where $\hat{a}^{\dagger}_{m}$ and $\hat{a}_{m}$ are the plasmon creation and annihilation operators, $\hat{\sigma}^{\dagger} $ and $\hat{\sigma} $ are the raising and lowering operators for the  QE, while the parameters $g$ and  $\bm{\mu}_{m}$ characterize, respectively, plasmon's coupling to the QE and EM field.

For plasmonic nanostructures with characteristic size smaller than the radiation wavelength, the coupling parameters can be obtained microscopically by relating them to system geometry and local field \cite{shahbazyan-prb21}. For such systems, the plasmon modes are determined by the quasistatic Gauss equation \cite{stockman-review}
$\bm{\nabla}\!\cdot\! \left [\varepsilon' (\omega_{m},\bm{r}) \bm{\nabla}\Phi_{m}(\bm{r})\right ]=0$,
where 
$\Phi_{m}(\bm{r})$ is the mode potential that defines the mode field $\bm{E}_{m}(\bm{r})=-\bm{\nabla}\Phi_{m}(\bm{r})$, which we choose to be real. To determine the plasmon dipole moment for optical transitions, we recast the Gauss's law as $\bm{\nabla}\!\cdot\! \left [\bm{E}_{m}(\bm{r})+4\pi \bm{P}_{m}(\bm{r})\right ]=0$,  where $\bm{P}_{m}(\bm{r})=\chi'(\omega_{m},\bm{r})\bm{E}_{m}(\bm{r})$ is  the electric polarization vector and $\chi=(\varepsilon-1)/4\pi$  is the plasmonic system susceptibility. The plasmon  dipole moment has the form
%
\begin{equation}
\label{mode-dipole}
\bm{p}_{m}=\int dV \bm{P}_{m}=\int dV \chi'(\omega_{m},\bm{r})\bm{E}_{m}(\bm{r}).
\end{equation}
The Gauss's equation  does not  determine the overall field normalization \cite{stockman-review}, but the later can be found  by matching the plasmon radiative decay rate and that of a localized dipole with excitation energy $\hbar \omega_{m}$. The plasmon radiative decay rate has the form \cite{shahbazyan-prb18} $\gamma_{m}^{r}=W_{m}^{r}/U_{m}$, where 
\begin{align}
\label{energy-mode}
U_{m}
= \frac{1}{16\pi} 
\!\int \!  dV \, \frac{\partial [\omega_{m}\varepsilon'(\omega_{m},\bm{r})]}{\partial \omega_{m}} \,\bm{E}_{m}^{2}(\bm{r}) ,
\end{align}
is the plasmon mode energy \cite{landau,shahbazyan-prl16} and 
\begin{equation}
\label{power}
W_{m}^{r}=\frac{p_{m}^{2}\omega_{m}^{4}}{3c^{3}},
\end{equation}
is the radiated power  ($c$ is the speed of light)  \cite{novotny-book}. The normalized modes $\tilde{\bm{E}}_{m}(\bm{r})$ are thus determined by setting 
\begin{equation}
\label{rate-mode-rad}
\gamma_{m}^{r}=\frac{4\mu_{m}^{2}\omega_{m}^{3}}{3\hbar c^{3}},
\end{equation}
where $\bm{\mu}_{m}$ is the mode optical transition matrix element. We then find the normalization relation as
\begin{equation}
\label{mode-rescaled}
\tilde{\bm{E}}_{m}(\bm{r})=\frac{1}{2}\sqrt{\frac{\hbar\omega_{m}}{U_{m}}} \bm{E}_{m}(\bm{r}),
\end{equation}
where the scaling factor $\sqrt{\hbar\omega_{m}/U_{m}}$ converts the plasmon  energy $U_{m}$  to $\hbar\omega_{m}$ in order to match the EM field energy (the factor $1/2$ reflects positive-frequency contribution). 
Accordingly, the plasmon optical transition matrix element in the Hamiltonian (\ref{H-full}) takes the form [compare to Eq.~(\ref{mode-dipole})]
\begin{equation}
\label{mode-optical}
\bm{\mu}_{m}=\int dV \chi'(\omega_{m},\bm{r})\tilde{\bm{E}}_{m}(\bm{r}).
\end{equation}
In a similar way, the plasmon non-radiative decay rate is  $\gamma_{m}^{nr}=W_{m}^{nr}/U_{m}$, where $W_{m}^{r}=\frac{1}{8\pi} 
\!\int \!  dV \varepsilon''(\omega_{m},\bm{r})\bm{E}_{m}^{2}(\bm{r})$ is the power dissipated in the plasmonic structure due to Ohmic losses. In terms of normalized fields, the non-radiative rate takes the form
\begin{equation}
\gamma_{m}^{nr}=\frac{1}{2\pi\hbar\omega_{m}} 
\!\int \!  dV \varepsilon''(\omega_{m},\bm{r})\tilde{\bm{E}}_{m}^{2}(\bm{r}),
\end{equation}
and so the plasmon full decay rate is $\gamma_{m}=\gamma_{m}^{nr}+\gamma_{m}^{r}$. Note that in structures with a single metallic component, the standard expression \cite{stockman-review} for $\gamma_{m}^{nr}$ is recovered: $\gamma_{m}^{nr}=2\varepsilon''(\omega_{m})/[\partial\varepsilon'(\omega_{m})/\partial \omega_{m}]$. The optical polarizability tensor of a plasmonic structure describing its response to the external field $\bm{E}e^{-i\omega t}$ has the form 
\begin{equation}
\label{polar-mode}
\bm{\alpha}_{m}(\omega)=\frac{1}{\hbar}\frac{\bm{\mu}_{m}\bm{\mu}_{m}}{\omega_{m}-\omega-\frac{i}{2}\gamma_{m}},
\end{equation}
where we kept only the resonance term \cite{shahbazyan-prb18}.

The QE-plasmon coupling in the Hamiltonian (\ref{H-full}) is expressed via normalized plasmon mode fields as \cite{shahbazyan-prb21}
\begin{equation}
\label{coupling}
\hbar g=-\bm{\mu}_{e}\!\cdot\!\tilde{\bm{E}}_{m}(\bm{r}_{e}).
\end{equation}
To present the coupling in a cavity-like form, we use the original plasmon mode fields (\ref{mode-rescaled}) to obtain \cite{shahbazyan-nl19}
\begin{equation}
\label{qe-pl-coupling-mode-volume}
g^{2} = \frac{2\pi \mu_{e}^{2}\omega_{m} }{\hbar {\cal V}},
~~~
\frac{1}{{\cal V}}  = \frac{2[\bm{n}_{e}\!\cdot\! \bm{E}_{m}(\bm{r}_{e})]^{2}}{\int \! dV [\partial (\omega_{m}\varepsilon')/\partial \omega_{m}]\bm{E}_{m}^{2}},
\end{equation}
where ${\cal V}$ is  projected plasmon mode volume characterizing   plasmon field confinement at the emitter position $\bm{r}_{e}$ along its dipole orientation $\bm{n}_{e}$ \cite{shahbazyan-prl16,shahbazyan-acsphot17,shahbazyan-prb18}. The plasmon mode volume defines the Purcell factor characterizing radiation enhancement  of a QE near a plasmonic structure:
\begin{equation}
F_{p}=\frac{\gamma_{e\rightarrow m}}{\gamma_{e}^{r}}=\frac{6\pi c^{3}Q_{m}}{\omega_{m}^{3} {\cal V}}
\end{equation}
where $Q_{m} =\omega_{m}/\gamma_{m}$ is  plasmon quality factor, $\gamma_{e}^{r}=4\mu_{e}^{2}\omega_{m}^{3}/3\hbar c^{3}$ is the emitter's radiative decay rate (at plasmon resonance frequency) and $\gamma_{e\rightarrow m}$ is the rate of energy transfer (ET) from QE to plasmon, given by
\begin{equation}
\label{qe-plasmon-rate-mode-volume}
\gamma_{e\rightarrow m}= \frac{8\pi \mu_{e}^{2}Q_{m} }{\hbar {\cal V}}.
\end{equation}
Comparing Eqs.~(\ref{qe-pl-coupling-mode-volume}) and (\ref{qe-plasmon-rate-mode-volume}), we obtain a relation between the QE-plasmon coupling and decay rates:
\begin{equation}
\label{coupling-rate}
g^{2}=\frac{1}{4}\gamma_{m}\gamma_{e\rightarrow m}=\frac{F_{p}}{4}\,\gamma_{m}\gamma_{e}^{r}.
\end{equation}
Thus, all coupling parameters in the Hamiltonian characterizing plasmon interactions with the QE and  EM field are expressed via  system parameters and related to plasmon and QE decay rates. Below, we employ these microscopic expressions to elucidate the role of ExIT and Fano interference in scattering spectra of hybrid plasmonic systems.

\section{Optical dipole moment of a hybrid plasmonic system}

We are interested in the linear response of hybrid plasmonic system to the external EM field. We assume that there is only a single excitation in the system and disregard any non-linear effects. In this case, we can approximate the QE by bosonic operators to setup Maxwell-Bloch (MB) equations for non-diagonal elements of density matrix (polarizations) $\rho_{e}(t)$ and $\rho_{m}(t)$ related to QE and plasmon induced dipoles as $\bm{p}_{e}(t)=\bm{\mu}_{e}\rho_{e}(t)$ and $\bm{p}_{m}(t)=\bm{\mu}_{m}\rho_{m}(t)$, respectively.  Using the Hamiltonian (\ref{H-full}), in the linear approximation, the MB equations for  $\rho_{m}(t)$  and $\rho_{e}(t)$ are obtained  in a standard manner as
\begin{align}
\label{mb}
&i\dot{\rho}_{m}=(\omega_{m}-i\gamma_{m}/2)\rho_{m}+g\rho_{e}- \bm{\mu}_{m}\!\cdot\!\bm{E}\, e^{-i\omega t},
\nonumber\\
&i\dot{\rho}_{e}=(\omega_{e}-i\gamma_{e}/2)\rho_{e}+g\rho_{m}- \bm{\mu}_{e}\!\cdot\!\bm{E}\, e^{-i\omega t},
\end{align}
where dot stands for the time-derivative and $\gamma_{e}$ is the QE spectral linewidth assumed much smaller than $\gamma_{m}$. 

In the steady-state case, substituting $\rho_{m}(t)=\rho_{m}e^{-i\omega t}$ and $\rho_{e}(t)=\rho_{e}e^{-i\omega t}$, we find
\begin{equation}
\rho_{m}=
\frac{\left (\omega_{e}-\omega - \frac{i}{2}\gamma_{e}\right )\bm{\mu}_{m}\!\cdot\!\bm{E} -g\bm{\mu}_{e}\!\cdot\!\bm{E}}
{\left (\omega_{m}-\omega -\frac{i}{2}\gamma_{m}\right )\!\!\left (\omega_{e}-\omega -\frac{i}{2}\gamma_{e}\right )-g^{2}} 
\end{equation}
and
\begin{equation}
\rho_{e}=
\frac{\left (\omega_{m}-\omega - \frac{i}{2}\gamma_{m}\right )\bm{\mu}_{e}\!\cdot\!\bm{E} -g\bm{\mu}_{m}\!\cdot\!\bm{E}}
{\left (\omega_{m}-\omega -\frac{i}{2}\gamma_{m}\right )\!\!\left (\omega_{e}-\omega -\frac{i}{2}\gamma_{e}\right )-g^{2}}.
\end{equation}
The system's  induced dipole moment is $\bm{p}_{s}=\bm{p}_{m}+\bm{p}_{e}=\bm{\mu}_{m}\rho_{m}+\bm{\mu}_{e}\rho_{e}$. To elucidate the processes contributing to $\bm{p}_{s}$, we define  QE polarizability tensor (in RWA) as
\begin{equation}
\label{polar-qe}
 \bm{\alpha}_{e}(\omega)=
 \frac{1}{\hbar}\frac{\bm{\mu}_{e}\bm{\mu}_{e}}{\omega_{e}-\omega-\frac{i}{2}\gamma_{e}},
 \end{equation} 
and introduce \textit{plasmon-induced} QE dipole moment as 
%
\begin{equation}
\label{qe-dipole-induced}
\bm{q}_{e}(\omega)=\bm{\alpha}_{e}(\omega)\tilde{\bm{E}}_{m}(\bm{r}_{e})=
\frac{\bm{\mu}_{e}}{\hbar}\frac{\bm{\mu}_{e}\!\cdot\!\tilde{\bm{E}}_{m}(\bm{r}_{e})}{\omega_{e}-\omega-\frac{i}{2}\gamma_{e}}.
\end{equation}
Then, the hybrid system dipole moment can be decomposed into three contributions:  
\begin{equation}
\bm{p}_{s}=\bm{p}_{dp}+\bm{p}_{int}+\bm{p}_{de}.
\end{equation}
The main contribution comes from the \textit{dressed plasmon} characterized by induced dipole moment 
\begin{equation}
\label{dipole-mode}
\bm{p}_{dp}=\frac{1}{\hbar}\frac{\bm{\mu}_{m}(\bm{\mu}_{m}\!\cdot\!\bm{E})}{\omega_{m}+\Sigma_{m}(\omega)-\omega-\frac{i}{2}\gamma_{m}},
\end{equation}
where
%
\begin{equation}
\label{self-mode}
\Sigma_{m}(\omega)=-\frac{g^{2}}{\omega_{e}-\omega-\frac{i}{2}\gamma_{e}}=-\bm{q}_{e}(\omega)\!\cdot\!\tilde{\bm{E}}_{m}(\bm{r}_{e}),
\end{equation}
is  plasmon's self-energy due to its interactions with the QE. Specifically, the imaginary part of self-energy determines the ET rate from the plasmon to QE as
\begin{equation}
\label{rate-me}
\gamma_{m\rightarrow e}(\omega)=-2\Sigma''_{m}(\omega)= \frac{ g^{2} \gamma_{e}}{(\omega-\omega_{e})^{2}+\gamma_{e}^{2}/4},
\end{equation}
which represents a Lorentzian centered at QE frequency $\omega_{e}$ and maximum value $\gamma_{m\rightarrow e}\equiv \gamma_{m\rightarrow e}(\omega_{e})=4g^{2}/\gamma_{e}$.

The QE-plasmon \textit{interference} term has the form
\begin{equation}
\label{dipole-mode-emitter}
\bm{p}_{int}=\frac{1}{\hbar}\frac{\bm{\mu}_{m}(\bm{q}_{e}\!\cdot\!\bm{E})+\bm{q}_{e}(\bm{\mu}_{m}\!\cdot\!\bm{E})}{\omega_{m}+\Sigma_{m}(\omega)-\omega-\frac{i}{2}\gamma_{m}},
\end{equation}
 and describes \textit{indirect}, i.e., mediated by plasmon, interactions of  QE with the EM field. The last term represents dressed QE contribution,
\begin{equation}
\label{dipole-emitter}
\bm{p}_{de}=\frac{1}{\hbar}\frac{\bm{\mu}_{e}(\bm{\mu}_{e}\!\cdot\!\bm{E})}{\omega_{e}+\Sigma_{e}(\omega)-\omega-\frac{i}{2}\gamma_{e}},
\end{equation}
where 
\begin{equation}
\label{self-emitter}
\Sigma_{e}(\omega)=-\frac{g^{2}}{\omega_{m}-\omega-\frac{i}{2}\gamma_{m}},
\end{equation}
is  the QE self-energy, whose imaginary part now determines the ET rate from the QE to plasmon as
\begin{equation}
\label{rate-em}
\gamma_{e\rightarrow m}(\omega)=-2\Sigma''_{e}(\omega)= \frac{ g^{2} \gamma_{m}}{(\omega-\omega_{m})^{2}+\gamma_{m}^{2}/4},
\end{equation}
which represents a Lorentzian centered at plasmon frequency $\omega_{m}$ and maximum value $\gamma_{e\rightarrow m}\equiv \gamma_{e\rightarrow m}(\omega_{m})=4g^{2}/\gamma_{m}$, matching Eq.~(\ref{coupling-rate}). Importantly,  in a narrow frequency interval  $|\omega-\omega_{e}|\lesssim \gamma_{e}$, the \textit{reverse} plamon-QE ET rate $\gamma_{m\rightarrow e}$  \textit{exceeds} the direct QE-plasmon ET rate $\gamma_{e\rightarrow m}$:
\begin{equation}
\label{rates-imbalance}
\frac{\gamma_{m\rightarrow e}}{\gamma_{e\rightarrow m}}= \frac{\gamma_{m}}{\gamma_{e}}\gg 1.
\end{equation}
Although the overall ET balance over the entire frequency range is preserved, the ET  \textit{imbalance} in the frequency interval $\sim \gamma_{e}$ leads to emergence of the ExIT minimum in the dressed plasmon spectra \cite{shahbazyan-prb20}.  For a typical case   $\mu_{e}/\mu_{m}\ll 1$, the dressed emitter's dipole moment (\ref{dipole-emitter}) is negligible small relative to dressed plasmon's dipole moment (\ref{dipole-mode}) and can be omitted. While the dressed plasmon approximation, i.e.,  $\bm{p}_{s}\approx\bm{p}_{dp}$, describes, with a reasonable accuracy, the position and magnitude of the ExIT minimum  in terms of energy exchange  between the QE and plasmon,   it does \textit{not} account for QE interactions with the EM field.  The latter is included \textit{indirectly} in the interference term (\ref{dipole-mode-emitter}) via plasmon-induced QE dipole moment $\bm{q}_{e}$, which, as we show below, gives rise to Fano interference that strongly affects the overall shape of scattering spectra as the system transitions to  strong coupling regime.

\section{Exciton-induced transparency vs. Fano interference}

The scattering cross-section  $\sigma_{s}^{sc}(\omega)$ of a hybrid plasmonic system is obtained  by normalizing the radiated power $W_{s}=(\omega^{4}/3c^{3})|\bm{p}_{s}(\omega)|^{2}$ with the incident flux $S=(c/8\pi)E^{2}$ \cite{novotny-book}. In the following, we disregard  the relatively small direct QE coupling with the EM field but include the indirect coupling via plasmon-induced dipole moment, so that the induced system dipole includes the interference term: $\bm{p}_{s}\approx\bm{p}_{dp}+\bm{p}_{int}$. The resulting expression for $\sigma_{s}^{sc}(\omega)$ is quite cumbersome as it depends sensitively on mutual polarizations of the incident light $\bm{E}$, the plasmon dipole moment $\bm{\mu}_{m}$ and the QE dipole moment $\bm{\mu}_{e}$. Here, to simplify the analysis, we consider the case when all dipole moments are parallel to the incident field, i.e. $\bm{\mu}_{e} \parallel \bm{\mu}_{m}\parallel \bm{E}$,  so that  the coupling between the system components and to the EM field is strongest. In this case, the two terms in the numerator in Eq.~(\ref{dipole-mode-emitter}) are equal, and using Eqs.~(\ref{qe-dipole-induced}) and (\ref{self-mode}), we obtain
\begin{equation}
\label{scattering}
\sigma_{s}^{sc}(\omega)=\frac{8\pi \omega^{4}}{3\hbar^{2}c^{4}}
\left|\frac{\mu_{m}^{2} \left (\omega_{e}+\omega_{F}-\omega - \frac{i}{2}\gamma_{e}\right )}
{\left (\omega_{m}-\omega -\frac{i}{2}\gamma_{m}\right )\!\!\left (\omega_{e}-\omega -\frac{i}{2}\gamma_{e}\right )-g^{2}}\right |^{2}, 
\end{equation}
 where $\omega_{F}=-2g\mu_{e}/\mu_{m}$ is QE frequency shift due to Fano interference between the plasmon  and plasmon-induced QE dipole moments as the system interacts with the EM field. In fact, this shift is the only difference between the current model and  dressed plasmon model (with $\bm{p}_{s}\approx\tilde{\bm{p}}_{m}$), which does not include the interference effects \cite{shahbazyan-prb20}.
To highlight the role of Fano interference, we relate the scattering cross-section (\ref{scattering}) to dressed plasmon scattering cross-section $ \sigma_{dp}^{sc}(\omega)$, which is obtained from (\ref{scattering}) by setting $\omega_{F}=0$, as
 \begin{equation}
 \sigma_{s}^{sc}(\omega)=\sigma_{dp}^{sc}(\omega)F(\omega),
 \end{equation}
 where $F(\omega)$ is the Fano function, 
\begin{equation}
\label{fano-function}
F(\omega)= \frac{(\delta-q)^{2}+1}{\delta^{2}+1}.
\end{equation}
Here,  $\delta=2(\omega-\omega_{e})/\gamma_{e}$ is frequency detuning in units of linewidth and $q$ is the Fano parameter:
\begin{equation}
\label{fanoQ}
q=\frac{2\omega_{F}}{\gamma_{e}}
=-\frac{4  g \mu_{e} }{\gamma_{e} \mu_{m}}.
\end{equation}
The  Fano function has  asymmetric shape that depends on the sign of parameter $q$. Using Eq.~(\ref{coupling-rate}), the magnitude of  $q$ can be expressed via the Purcell factor as
\begin{equation}
\label{fanoQ2}
|q|= \frac{2\gamma_{e}^{r}}{\gamma_{e}} \, \sqrt{\frac{F_{p}}{\eta_{m}}},
\end{equation}
where $\eta_{m}=\gamma_{m}^{r}/\gamma_{m}$ is the plasmon radiation efficiency. Although the ratio $\gamma_{e}^{r}/\gamma_{e}$ is typically very small ($\sim 10^{-5}$) due to  the broadening of spectral linewidth $\gamma_{e}$ by phonons or vibrons, for small nanostructures  we have $F_{p}\gg 1$ and $\eta_{m}\ll 1$, implying that, in a plasmonic hot spot, the actual value of  $q$  can be appreciable.

To elucidate the interplay between Fano interference and ExIT, we recall that, in the scattering spectra, the ExIT minimum emerges in the weak coupling regime as a narrow dip on the top of a wide plasmon band. The plasmon scattering cross-section is obtained by setting $g=0$ in Eq.~(\ref{scattering}) and, for $\bm{\mu}_{m}\parallel \bm{E}$, has the form
\begin{equation}
\label{scattering-mode}
\sigma_{m}^{sc}(\omega)=
\frac{8\pi \omega^{4}}{3\hbar^{2}c^{4}}\frac{\mu_{m}^{4}}{(\omega_{m}-\omega)^{2}+\gamma_{m}^{2}/4}.
\end{equation}
To trace the emergence of ExIT minimum, we recast the dressed plasmon scattering cross-section as $\sigma_{dp}^{sc}(\omega)=\sigma_{m}^{sc}(\omega)R(\omega)$, where the function
\begin{equation}
\label{R-full}
R(\omega)=\left |\frac{\left (\omega_{m}-\omega-\frac{i}{2}\gamma_{m}\right )\left (\omega_{e}-\omega-\frac{i}{2}\gamma_{e}\right )}
{\left (\omega_{m}-\omega-\frac{i}{2}\gamma_{m}\right )\left (\omega_{e}-\omega-\frac{i}{2}\gamma_{e}\right )-g^{2}}\right |^{2}
\end{equation}
modulates the plasmon band, and so the system scattering cross-section is factorized as 
\begin{equation}
\label{sigma-sc}
\sigma_{\rm s}^{\rm sc}(\omega)=\sigma_{m}^{\rm sc}(\omega)R(\omega)F(\omega).
\end{equation}
In the frequency interval $|\omega_{m}-\omega|/\gamma_{m}\ll 1$, using the relation (\ref{coupling-rate}), the function  $R(\omega)$  simplifies to
\begin{equation}
\label{R-weak}
R(\omega)=\frac{\delta^{2}+1}
{\delta^{2}+(1+p)^{2}},
\end{equation}
where the parameter  
\begin{equation}
\label{exitP}
p=\frac{\gamma_{e\rightarrow m}}{\gamma_{e}}=\frac{4g^{2}}{\gamma_{m}\gamma_{e}}
\end{equation}
characterizes the  ExIT minimum depth. The ExIT function (\ref{R-weak}) describes the emergence of spectral minimum  due to excessively large plasmon-QE ET in the frequency interval $\sim \gamma_{e}$. Specifically, in the weak coupling regime, the  dressed plasmon decay rate has the form $\gamma_{dp}(\omega)=\gamma_{m}+\gamma_{m\rightarrow e}(\omega)$.  Using Eq.~(\ref{rate-me}) and the relation (\ref{coupling-rate}), we obtain
\begin{equation}
\label{plasmon-rate-qe}
\gamma_{dp}(\omega)
=\gamma_{m}\left (1+ \frac{ p}{\delta^{2}+1}\right ),
\end{equation}
implying linewidth increase by  factor $(1+p)$  in  the frequency interval $|\omega-\omega_{e}|\sim \gamma_{e}$ which, in turn,  leads to the ExIT minimum in the dressed plasmon spectrum. 
 
Thus, in the weak coupling regime, the ExIT and Fano interference effects are distinct and described by different factors in the scattering cross-section (\ref{sigma-sc}). While the ExIT factor $R(\omega)$ leads to a narrow minimum at the QE frequency position, the Fano factor $F(\omega)$ is an asymmetric function of $\omega$ that affects the overall shape of the scattering spectra. Remarkably, as we show in numerical calculations below, the Fano interference effect is most visible  for intermediate and strong QE-plasmon coupling as it shifts the spectral weight between polaritonic bands  resulting in the \textit{inversion} of spectral asymmetry.

\section{Numerical results and discussion}

In this section, we present the results of numerical calculations for a QE situated at a distance $d$ from the tip of an Au nanorod in water with excitation frequency in resonance with the surface plasmon frequency, $\omega_{e}=\omega_{m}$. The nanorod was modeled by a prolate spheroid with semi-major and semi-minor axes $a$ and $b$, respectively, the QE's dipole orientation was chosen along the nanorod symmetry axis,  the Au experimental dielectric function  was used in all calculations \cite{johnson-christy}, and the dielectric constant of water was taken as $\varepsilon_{s}=1.77$. We used the standard spherical harmonics for calculations of the local fields near prolate spheroid  to obtain the plasmon parameters  $\mu_{m}$, $\gamma_{m}$, $\eta_{m}$, the QE-plasmon coupling $g$ and the  Purcell factor $F_{p}$, which determine the ExIT parameter $p$ and Fano parameter $q$. The QE spectral linewidth $\gamma_{e}$ was chosen much smaller than the plasmon decay rate, $\gamma_{e}/\gamma_{m}=0.1$, and its radiative decay time was chosen $\tau_{e}^{r}= 10$ ns, which are typical values for excitons in  quantum dots. Note that the  QE radiative decay rate $\gamma_{e}^{r}$ is much smaller that its spectral linewidth: for our system we have $\gamma_{e}^{r}/\gamma_{e}\sim 10^{-5}$.

%
\begin{figure}[tb]
\begin{center}
\includegraphics[width=0.8\columnwidth]{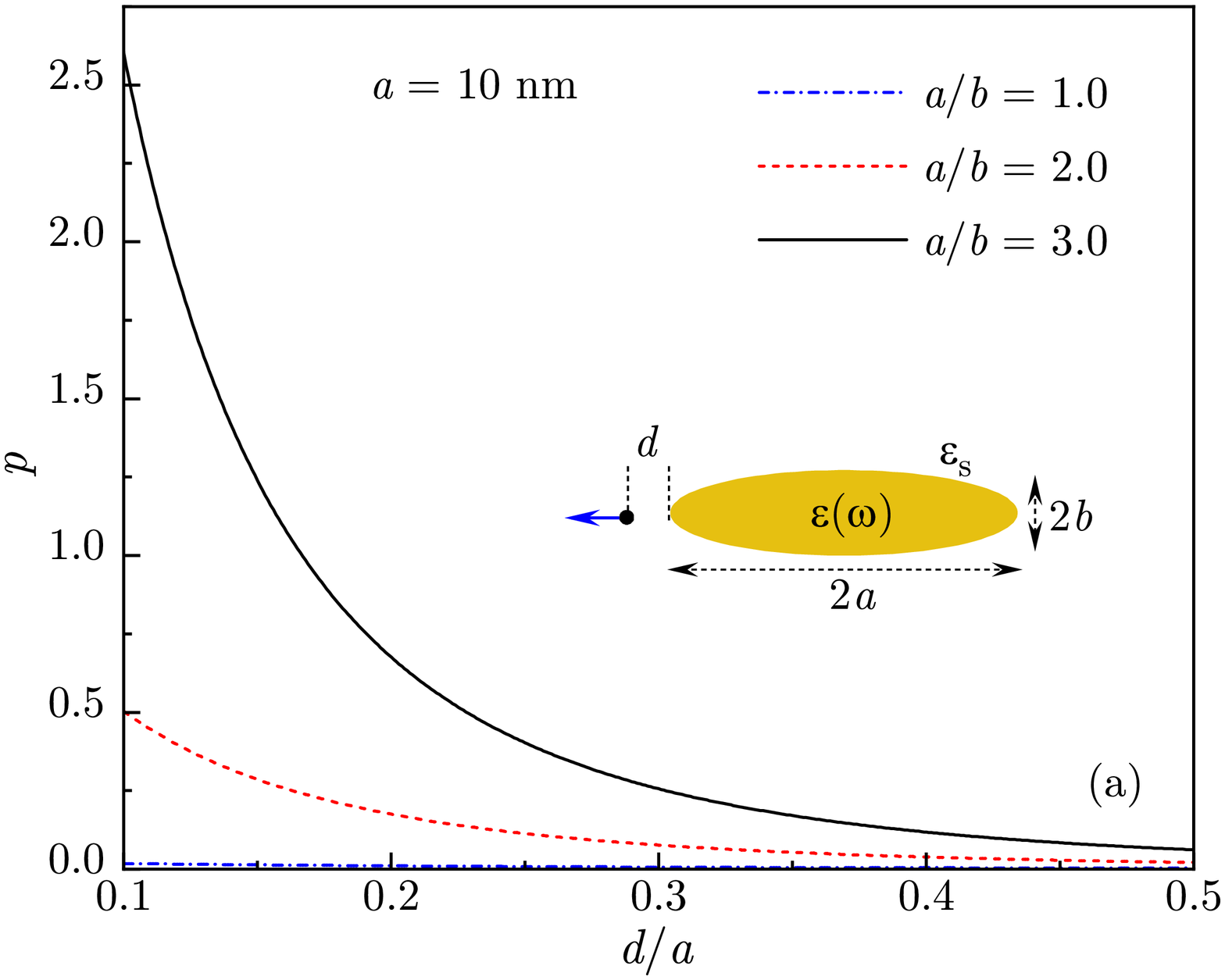}

\vspace{3mm}

\includegraphics[width=0.8\columnwidth]{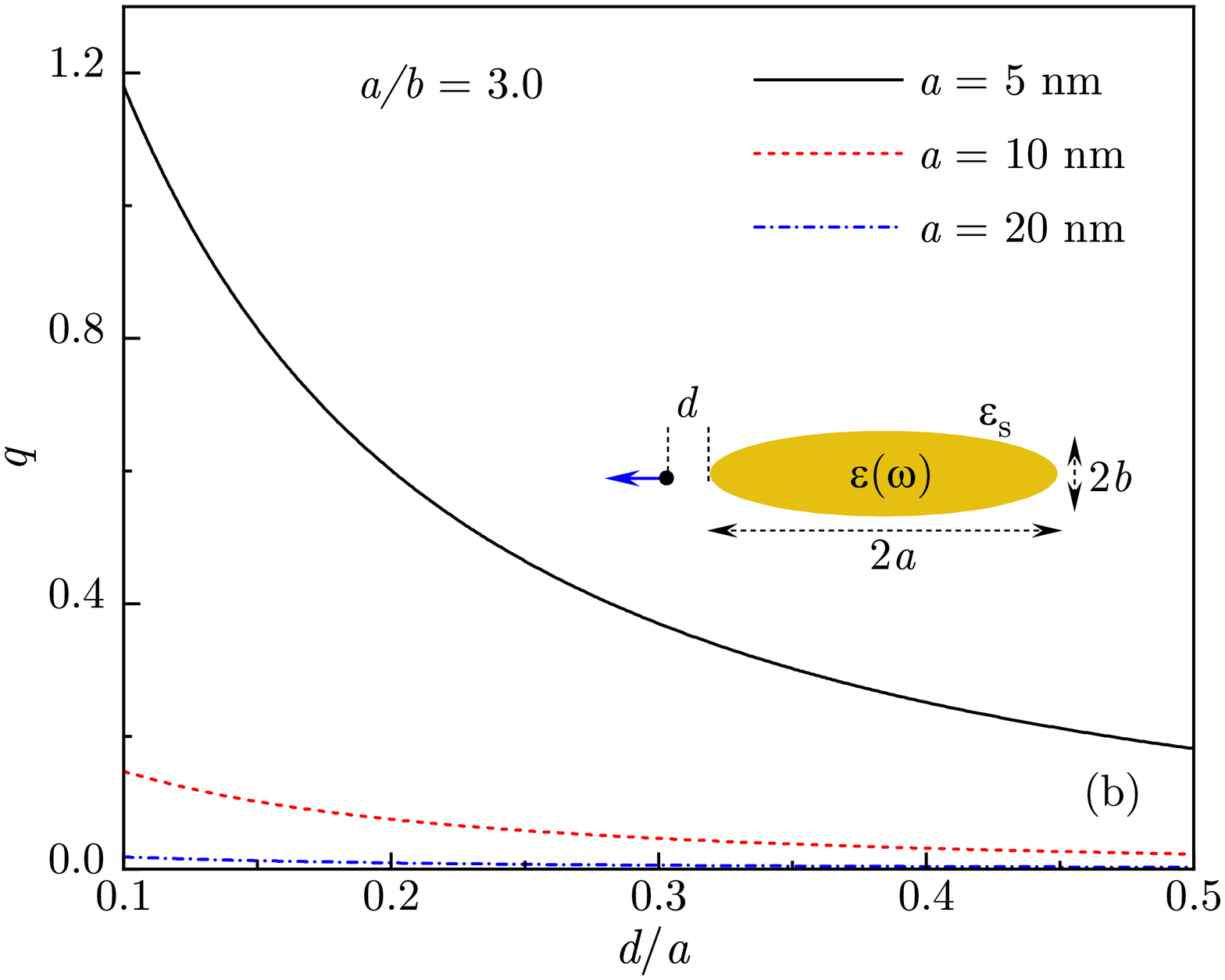}
\caption{\label{fig1} (a) The ExIT parameter $p$ is plotted against the QE distance $d$ to the tip of Au nanorod of length $2a=20$ nm placed for different values of aspect ratio $a/b=1.0$, 2.0 and 3.0.  (b) The Fano parameter $q$  is plotted against  the distance $d$ at nanorod aspect ratio $a/b=3.0$  for different values of nanorod  length $2a=40$ nm, 20 nm and 10 nm. Inset: Schematics of a QE situated at a distance $d$ from the tip of Au nanorod in water for QE dipole moment oriented along the nanorod axis.
 }
\end{center}
\end{figure}
%

In Fig.~\ref{fig1}, we plot  the calculated ExIT parameter $p$, given by Eq.~(\ref{exitP}), and the Fano parameter $q$, given by Eq.~(\ref{fanoQ}), against the distance to nanorod tip $d$ normalized by  $a$. Fig.~\ref{fig1}(a) shows the ExIT parameter $p=F_{p}\gamma_{e}^{r}/\gamma_{e}$ for three different values of nanorod aspect ratio: $a/b=1.0$ (sphere), 2.0 and 3.0. Note that the Purcell factor near the tip of elongated particle ($a/b=3.0$) is much greater that for a nanosphere ($a/b=1$),  so that $p> 1$ in the former case while being negligibly small in the latter case.  In Fig.~\ref{fig1}(b), we show distance dependence of the Fano parameter $q$ for fixed nanorod aspect ratio $a/b=3.0$ and different lengths $2a=40$ nm, 20 nm and 10 nm.  The Fano parameter is largest for the smallest nanorod with $2a=10$ nm and is significantly reduced for larger nanorods with $2a=20$ nm and 40 nm. Both $p$ and $q$ sharply decrease as the QE moves away from the hot spot near nanorod tip.

%
\begin{figure}[tb]
\begin{center}
\includegraphics[width=0.8\columnwidth]{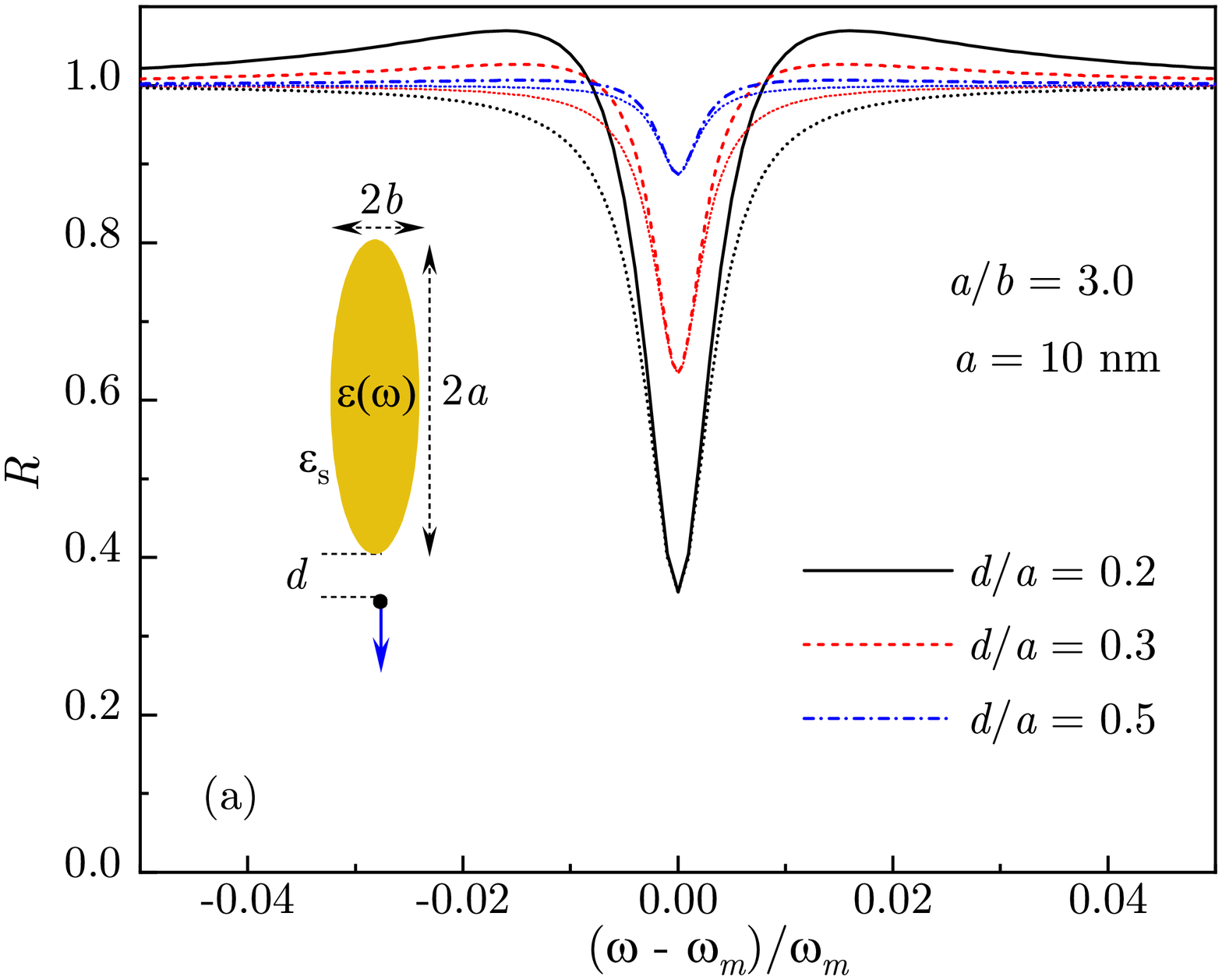}

\vspace{3mm}

\includegraphics[width=0.8\columnwidth]{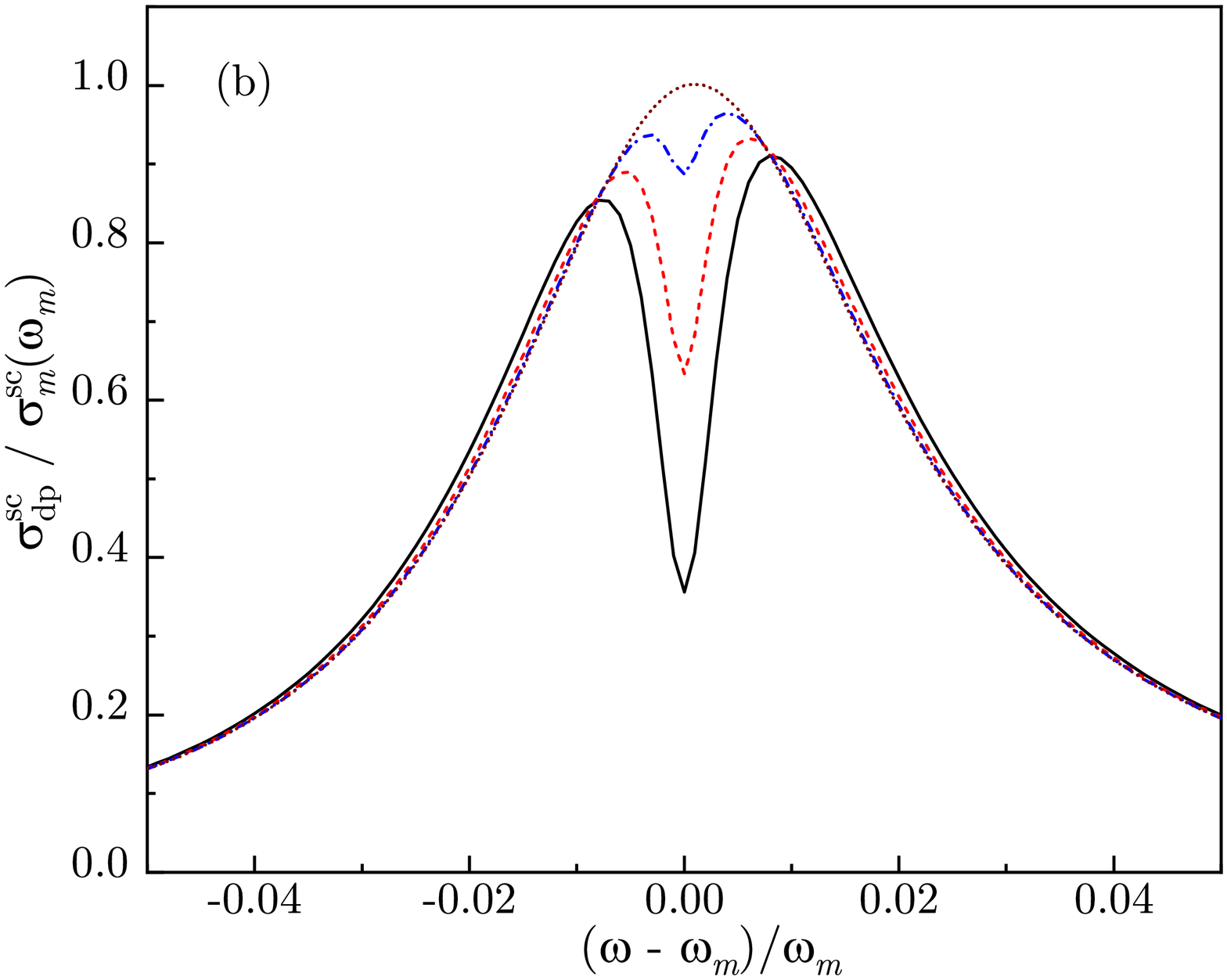}
\caption{\label{fig2} (a) The ExIT function $R(\omega)$, given by Eq.~(\ref{R-full}), and its asymptotic expression (dotted lines), given by Eq.~(\ref{R-weak}), are shown  for a QE near the tip of Au nanorod  with aspect ratio $a/b=3.0$ and length $2a=20$ nm at distances $d/a=0.5$, 0.3, and 0.2. (b)  Normalized scattering cross-section in the dressed plasmon approximation is shown for the same system parameters. Dotted line is the plasmon band. All curves are calculated for $\omega_{e}=\omega_{m}$. Inset: Schematics of a QE situated at a distance $d$ from the tip of Au nanorod in water for QE dipole moment oriented along the nanorod axis.
 }
\end{center}
\end{figure}
%

%
\begin{figure}[tb]
\begin{center}
\includegraphics[width=0.8\columnwidth]{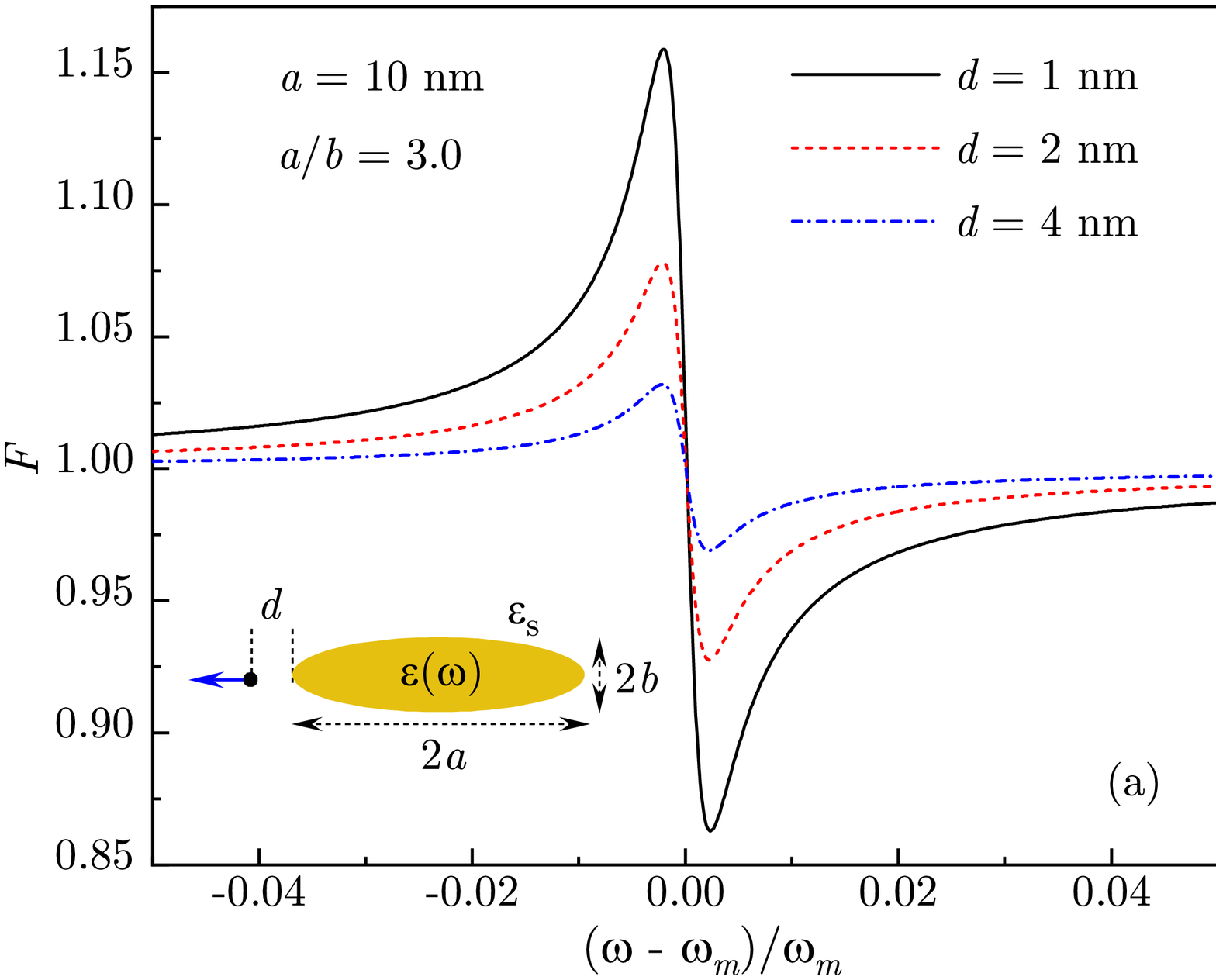}

\vspace{3mm}

\includegraphics[width=0.8\columnwidth]{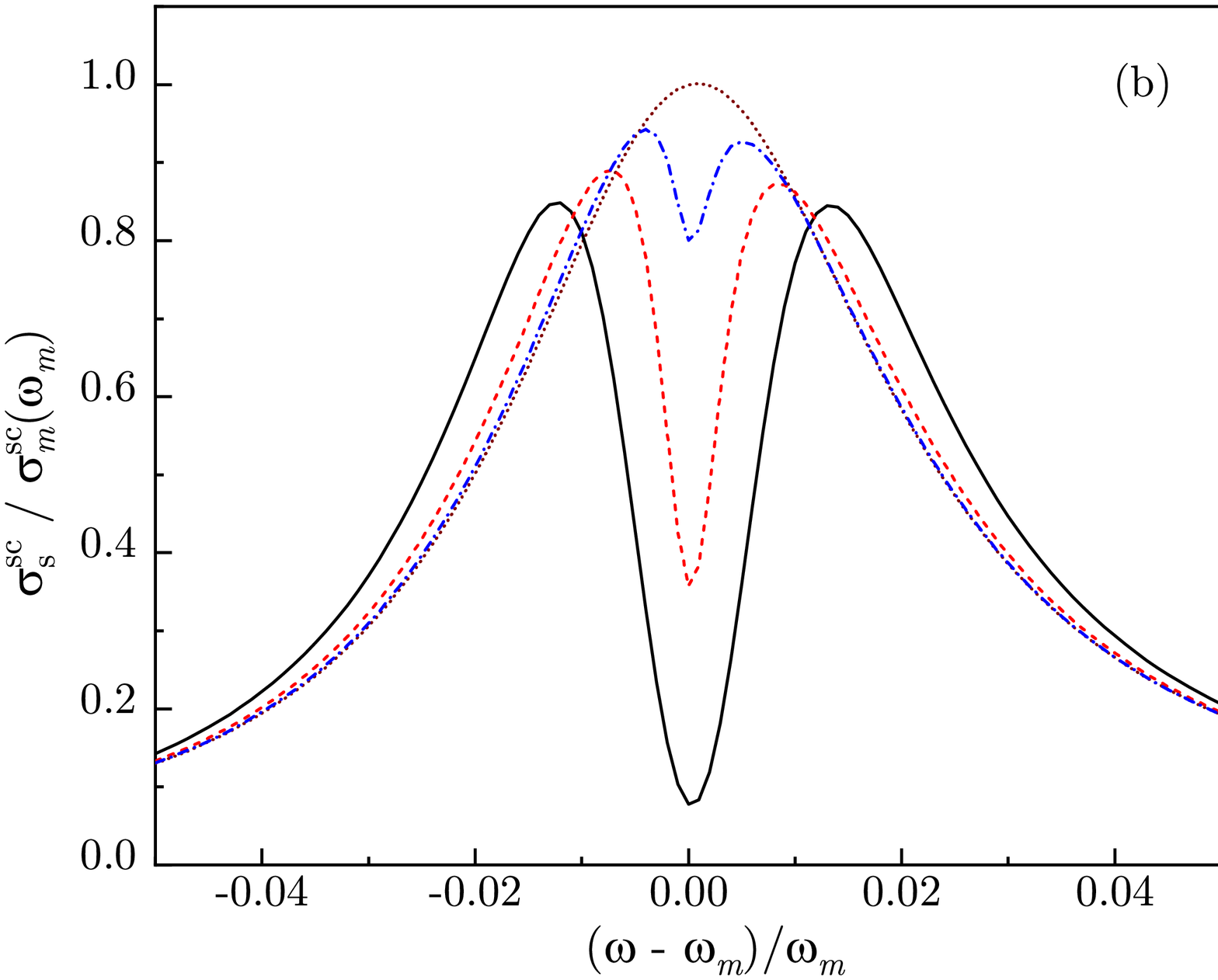}
\caption{\label{fig3} (a) The Fano function $F(\omega)$, given by Eq.~(\ref{fano-function}), is shown  for a QE near the tip of Au nanorod  with aspect ratio $a/b=3.0$ and length $2a=20$ nm at distances $d/a=0.5$, 0.3, and 0.2. (b)  Normalized scattering cross-section, given by Eq.~(\ref{sigma-sc}),  is shown for the same system parameters. Dotted line is the plasmon band. All curves are calculated for $\omega_{e}=\omega_{m}$. Inset: Schematics of a QE situated at a distance $d$ from the tip of Au nanorod in water for QE dipole moment oriented along the nanorod axis.
 }
\end{center}
\end{figure}
%

In Fig.~\ref{fig2}, in order to illustrate the emergence of ExIT,  we show the evolution of function $R(\omega)$, given by Eq.~(\ref{R-full}), and of dressed plasmon's scattering cross-section $\sigma_{dp}^{\rm sc}(\omega)=\sigma_{m}^{\rm sc}(\omega)R(\omega)$ (i.e., without Fano interference effect) with decreasing QE-nanorod distance for $2a=20$ nm and $a/b=3.0$. With decreasing  $d$, the function $R(\omega)$ develops a minimum, as shown in Fig.~\ref{fig2}(a), which modulates the plasmon scattering spectrum, as shown in Fig.~\ref{fig2}(b).  as discussed in the previous section, the double-peak structure of $\sigma_{\rm dp}^{\rm sc}(\omega$ is caused by ET imbalance between the QE and plasmon in a narrow frequency interval.  In order to highlight the role of ExIT parameter $p$,  we plot in Fig.~\ref{fig2}(a) the asymptotic expression  for $R(\omega)$, given by Eq.~(\ref{R-weak}), for each value of QE-nanorod distance $d$ (dotted lines). Clearly, in the weak coupling regime (small $p$), the ExIT function Eq.~(\ref{R-weak}) accurately describes the spectral minimum (blue curves), while for larger $p$ (i.e., closer to the tip) the spectrum develops "wings" outside the minimum region as the system undergoes strong coupling transition. The onset of strong coupling transition can be seen in  Fig.~\ref{fig2}(b) as well, as for $d/a=0.5$ and 0.3, the scattering spectrum develops a narrow ExIT minimum at  QE frequency on top of unchanged plasmon band, while for $d/a=0.2$, the overall spectral width slightly increases signaling the emergence of Rabi splitting. We note that the ExIT function Eq.~(\ref{R-weak}) accurately reproduces the central part of  ExIT minimum for any distance $d$.

While the dressed plasmon model describes the position and depth of ExIT minimum relatively well, it predicts a sustained asymmetry as the spectral weight as the higher frequency region of scattering spectrum carries a larger spectral weight  [see Fig.~\ref{fig2}(b)]. In the absence of QE coupling to the EM field, emission takes place from the plasmonic antenna, whose power spectrum is $\propto \omega^{4}$ due to larger radiation rate at higher frequencies. Therefore, in the presence of double peak structure due to either  ExIT minimum or Rabi splitting centered at resonance frequency $\omega=\omega_{m}=\omega_{e}$, the higher frequency peak is enhanced. Note that similar scattering spectra are predicted by the classical model of coupled oscillators which disregards optical interference effects \cite{pelton-oe10,pelton-nc18}. Below we demonstrate that extending the dressed plasmon model to include Fano interference between the plasmon antenna and plasmon-induced QE dipole, as described in Eq.~(\ref{sigma-sc}), can strongly affect the overall shape of scattering spectra.

%
\begin{figure}[tb]
\begin{center}
\includegraphics[width=0.8\columnwidth]{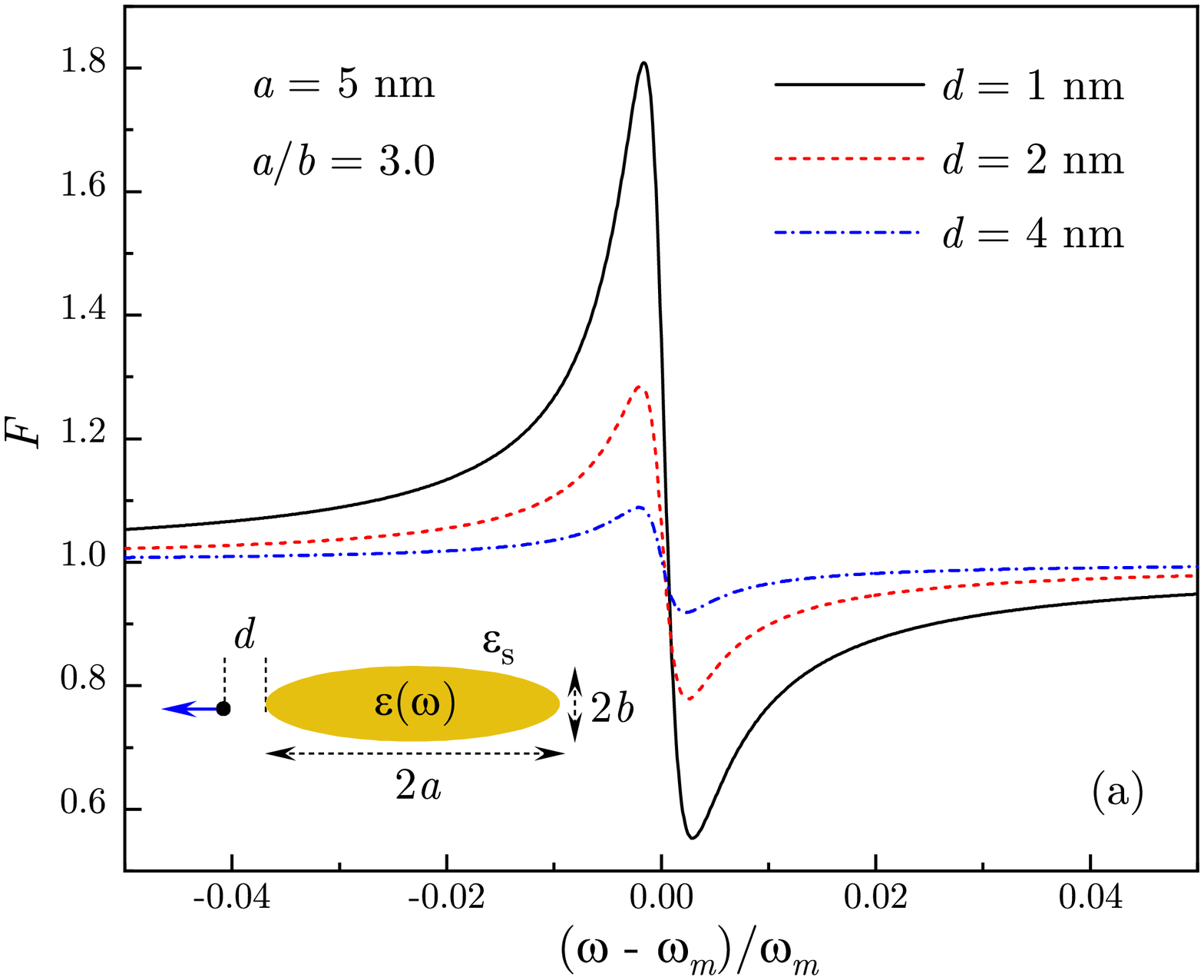}

\vspace{3mm}

\includegraphics[width=0.8\columnwidth]{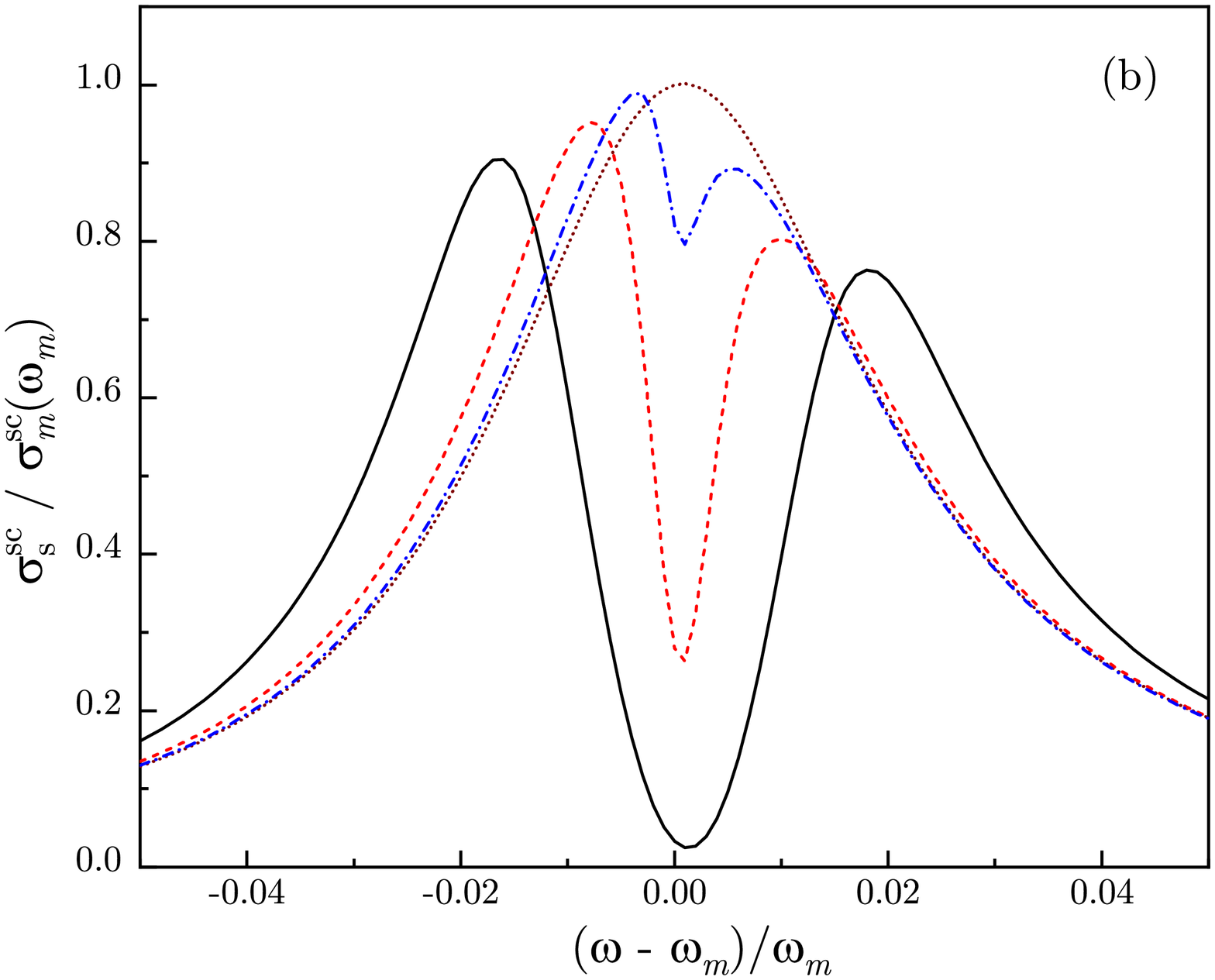}
\caption{\label{fig4} (a) The Fano function $F(\omega)$, given by Eq.~(\ref{fano-function}), is shown  for a QE near the tip of Au nanorod  with aspect ratio $a/b=3.0$ and length $2a=10$ nm at distances $d/a=0.5$, 0.3, and 0.2. (b)  Normalized scattering cross-section, given by Eq.~(\ref{sigma-sc}),  is shown for the same system parameters. Dotted line is the plasmon band. All curves are calculated for $\omega_{e}=\omega_{m}$. Inset: Schematics of a QE situated at a distance $d$ from the tip of Au nanorod in water for QE dipole moment oriented along the nanorod axis.
 }
\end{center}
\end{figure}
%

In Fig.~\ref{fig3}, we plot the Fano function and scattering spectra for a QE situated at several distances from the tip of Au nanorod with aspect ratio $a/b=3.0$ and overall length $2a=20$ nm. As indicated above, we consider the case of QE's dipole moment oriented along the normal to tip surface (see inset in Fig.~\ref{fig3}), so that the Fano parameter $q$ is positive.  For nanorod of this length, $q$ is relative small [see Fig.~\ref{fig1}(b)] and so the Fano function's variation ranges from about  2\% for $d=4$ nm to 15\% for  $d=1$ nm, as the QE-plasmon coupling $g$ increases close to the tip [see Fig.~\ref{fig3}(a)]. Importantly, for $q>0$, the  spectral shape of the Fano function, which enters in the scattering cross-section (\ref{sigma-sc}), leads to the suppression of higher frequency region and enhancement of lower frequency region. As a result, the aforementioned asymmetry of dressed plasmon scattering spectra in Fig.~\ref{fig2}(b) is largely compensated, and so the full scattering spectra are now close to symmetric  [see Fig.~\ref{fig3}(b)].

In Fig.~\ref{fig4}, we show the Fano function and scattering spectra for a small nanorod of length 10 nm. With decreasing nanostructure size and, hence, the reduction of plasmon mode volume, the QE-plasmon coupling increases  and so does the Fano parameter $q$, which now reaches values $q \sim 1$ [see Fig.~\ref{fig1}(b)]. In this case, the  Fano function variation is larger as well, reaching about 80\%  close to the nanorod tip [see Fig.~\ref{fig4}(a)]. As a result, the scattering  spectra, shown in Fig.~\ref{fig4}(b), exhibit \textit{inversion} of spectral asymmetry relative to the dressed plasmon spectra [see Fig.~\ref{fig2}(b)], with the lower frequency peak now substantially higher that the higher frequency peak. Note that for smallest $d$, the system has clearly transitioned to strong coupling regime since the double-peak structure is well beyond the plasmon resonance envelope. We stress that although the mechanisms of ExIT and Fano interference are distinct, as discussed in the previous section, the two effects are intimately related as  Fano interference maifests itself via redistribution of spectral weight across the ExiT minimum  in the scattering spectra.

\section{Conclusions}
In this paper, we developed a model for exciton-induced transparency (ExIT) and Fano interference in hybrid plasmonic systems comprised of a single emitter resonantly coupled to a surface plasmon in a metal-dielectric structure. We have shown that the  shape of scattering spectra is determined by   two distinct mechanisms. First is near-field coupling between the emitter and plasmon that defines the energy spectrum of hybrid system. This mechanism relies upon energy exchange between the system components  and gives rise to the ExIT minimum in scattering spectra and, in the strong coupling regime, to the Rabi splitting of polaritonic bands. The second mechanism is the Fano interference between the plasmon and the plasmon-induced emitter's dipoles as the system interacts with the radiation field. Although the Fano interference does not significantly affect the position or magnitude of ExIT minimum, it determines the overall shape of scattering spectra. Specifically, the Fano interference leads to the inversion of spectral asymmetry that was recently reported in the experiment \cite{zhang-nl17,xu-nl17,shegai-acsphot19}.


This work was supported in part by the National Science Foundation Grant Nos.  DMR-2000170, DMR-1856515 and    DMR-1826886.

\clearpage

\end{document}